\documentclass[10pt,superscriptaddress,aps,prx,twocolumn,showpacs,nofootinbib, longbibliography, notitlepage,floatfix]{revtex4-2}
\usepackage{amsmath,amssymb,amsthm}
\usepackage[pdftex]{graphicx}
\usepackage{times,txfonts}
\usepackage{braket}
\usepackage{color}
\usepackage{natbib}
\usepackage{amsmath,blkarray}
\usepackage{mathtools}
\usepackage[normalem]{ulem}
\usepackage{latexsym}
\usepackage{tabularx, booktabs}
\usepackage{graphics,epstopdf}
\usepackage{graphicx}
\usepackage{amsfonts}
\usepackage{color,soul}
\usepackage{multirow}
\usepackage{array}
\usepackage{appendix}
\usepackage{url}
\usepackage[colorlinks=true,citecolor=blue,urlcolor=blue]{hyperref}

\newcommand{\be}{\begin{equation}}
\newcommand{\ee}{\end{equation}}
\newcommand{\ba}{\begin{eqnarray}}
\newcommand{\ea}{\end{eqnarray}}

\begin{document}
\title{On the performance of nanoscale detection methods in NV-center spectroscopy }
\author{Koushik Paul}
\email{koushikpal09@gmail.com}
\affiliation{Department of Physical Chemistry, University of the Basque Country UPV/EHU, Apartado 644, 48080 Bilbao, Spain}
\affiliation{EHU Quantum Center, University of the Basque Country UPV/EHU, Barrio Sarriena, s/n, 48940 Leioa, Spain}
\author{Pol Alsina-Bolívar}
\affiliation{Department of Physical Chemistry, University of the Basque Country UPV/EHU, Apartado 644, 48080 Bilbao, Spain}
\affiliation{EHU Quantum Center, University of the Basque Country UPV/EHU, Barrio Sarriena, s/n, 48940 Leioa, Spain}
\author{Jorge Casanova}
\affiliation{Department of Physical Chemistry, University of the Basque Country UPV/EHU, Apartado 644, 48080 Bilbao, Spain}
\affiliation{EHU Quantum Center, University of the Basque Country UPV/EHU, Barrio Sarriena, s/n, 48940 Leioa, Spain}
\begin{abstract}
Nuclear magnetic resonance (NMR) in subfemtoliter volumes operates in a regime where statistical fluctuations of the nuclear magnetization can exceed its thermal mean. Solid-state quantum sensors such as nitrogen-vacancy (NV) centers in diamond are sensitive enough to detect NMR signals from such small volumes, which has motivated several detection protocols, based either on correlating the statistical polarization or on coherently detecting hyperpolarized samples. Here we bring Qdyne, CS, MCS, and CASR into a single framework for nanoscale samples. Specifically, we derive a unified analytical expression for the SNR and spin-number sensitivity applicable to all four protocols under consistent experimental conditions. Numerical simulations of the full measurement sequences confirm this expression over the whole range of parameters explored. This formulation enables a quantitative comparison of their performance while accounting for the distinct measurement sequences, acquisition times, and other related parameters. Using the SNR and the spin-number sensitivity as figures of merit, we identify which protocol performs best in each experimental regime, thereby providing a practical guide for choosing and tuning the most suitable method for a given experiment.
\end{abstract}
\maketitle
\section{Introduction}\label{Intro}
Determining the structure of single molecules, analyzing surfaces, and performing material analysis require nanoscale spatial resolution, typically tens of nanometers. Reaching this resolution would open the route to nanoscale nuclear magnetic imaging~\cite{budakian2024roadmap,bruckmaier2023imaging,meriles2010imaging,maletinsky2012robust}. Such capabilities are of central interest in fields ranging from chemistry and biology to materials science. Meeting these needs requires detecting the NMR signal from extremely small spin ensembles of $N \sim 10^{3}$--$10^{5}$ nuclear spins, contained in sample volumes of roughly $(3$--$10~\mathrm{nm})^{3}$, many orders of magnitude smaller than the volumes of at least ${\sim}1~\mathrm{mm}^{3}$ required by conventional inductive detection. A leading approach to overcome this limitation relies on solid-state quantum sensors, such as the nitrogen-vacancy (NV) center in diamond~\cite{allert2022advances,doherty2013nitrogen,rizzato2023quantum}, whose proximity to the sample provide a far higher volume sensitivity and has enabled experimental detection of nanoscale samples~\cite{du2024single,allert2022advances,rizzato2023quantum,degen2017quantum,schwartz2019blueprint}, both with single NVs~\cite{aslam2017nanoscale,mamin2013nanoscale,staudacher2013nuclear,lovchinsky2016nuclear,lovchinsky2017magnetic,abendroth2022single} and with ensembles~\cite{devience2015nanoscale,liu2022surface,ishiwata2021label,kehayias2017solution,ziem2019quantitative,martin2026coherent}.

At the nanoscale, the small number of nuclear spins contributing to the detected signal makes the statistical fluctuations of the polarization, which scale as $P_{\rm stat}\propto N^{-1/2}$~\cite{meriles2010imaging,herzog2014boundary,tabatabaei2024large}, several orders of magnitude larger than the mean (thermal) polarization, $P_{\rm th}=\hbar\gamma B/(2k_BT)$. It is therefore advantageous either to detect these statistical fluctuations, typically by correlating successive measurements, or to enhance the mean polarization by hyperpolarization.

Among many existing detection methods, we focus on the main phase-sensitive ones that can be applied in the nanoscale setting. Specifically, we inspect: 1) Quantum heterodyne detection (Qdyne)~\cite{schmitt2017submillihertz,boss2017quantum}, where the sensors sense the phase of the signal at successive times and the fluorescence time-trace is recorded. Each time trace is autocorrelated separately in post-processing. As a consequence, an NV ensemble only improves the performance of Qdyne over a single NV if the fluorescence of each NV is read out individually~\cite{cambria2025scalable,cheng2025massively}. 2) Correlation Spectroscopy (CS)~\cite{laraoui2013high,rosskopf2017quantum} performs the correlation directly within the quantum state of the NV sensor, eliminating the need for post-processing. However, the required long idle times reduce the SNR. 3) In Multipoint Correlation Spectroscopy (MCS)~\cite{spohn2025quantum,maier2025efficient} the full time trace is acquired within a single run while performing  phase correlations within the quantum state of the sensor with a reference phase stored in a quantum memory, here the nitrogen nuclear spin of the NV. 4) Alternatively, the mean contribution of the polarization itself can be detected after enhancing it with hyperpolarization techniques~\cite{bucher2020hyperpolarization,arunkumar2021micron}. This approach allows to employ usual coherent sensing sequences, such as Coherently Averaged Synchronized Readout (CASR)~\cite{glenn2018high}, offering higher spectral resolution and comparable or higher SNR~\cite{tabatabaei2024large,von2025coherent} to those obtained with correlation techniques. 

To reconstruct the time trace of the nuclear signal, every protocol must (i) target either the statistical or the mean contribution of the sample polarization, (ii) sample the signal with successive phase measurements, and (iii) turn these measurements into a time trace that is reproducible from run to run, so that it can be averaged. The protocols therefore differ in \emph{what} they target, \emph{how} they sample the signal, and \emph{when} they remove the random phase of the statistical contribution. Each of these choices sets one dependence of the SNR and sensitivity: the target sets its dependence on the sample polarization, the sampling sets the cost of the frequency resolution, and the moment of the correlation sets how the readout noise enters. As a result, no single method is optimal in every scenario~\cite{staudenmaier2023optimal,spohn2025quantum,von2025coherent}, and choosing among them is not straightforward. 

Here, we unify Qdyne, CS, MCS, and CASR within a single theoretical framework. Our central result is a single analytical expression for the SNR, and from it the spin-number sensitivity, that covers all four protocols under the same experimental conditions. Numerical simulations of the measurement sequences confirm this expression in a wide parameter regime. With this, we trace the performance of each protocol back to the \textit{what}, \textit{how}, and \textit{when} distinctions introduced above, and identify the best protocol for each experimental regime. 

The rest of the paper is organized as follows. In Sec.~\ref{sec:2}, we introduce the four detection protocols and establish a common framework for their signal acquisition and measurement noise. In Sec.~\ref{sec:snr}, we derive the analytical expressions for their SNRs, accounting for the relevant experimental parameters and coherence times. In Sec.~\ref{sec:4} and \ref{sec:5}, we present a systematic comparison of the protocols, examining the effects of readout noise, coherence time, and spectral resolution. Finally, we summarise our main findings and conclude in Sec.~\ref{sec:conc}.
\begin{figure*}
    \centering
\includegraphics[width=1\linewidth]{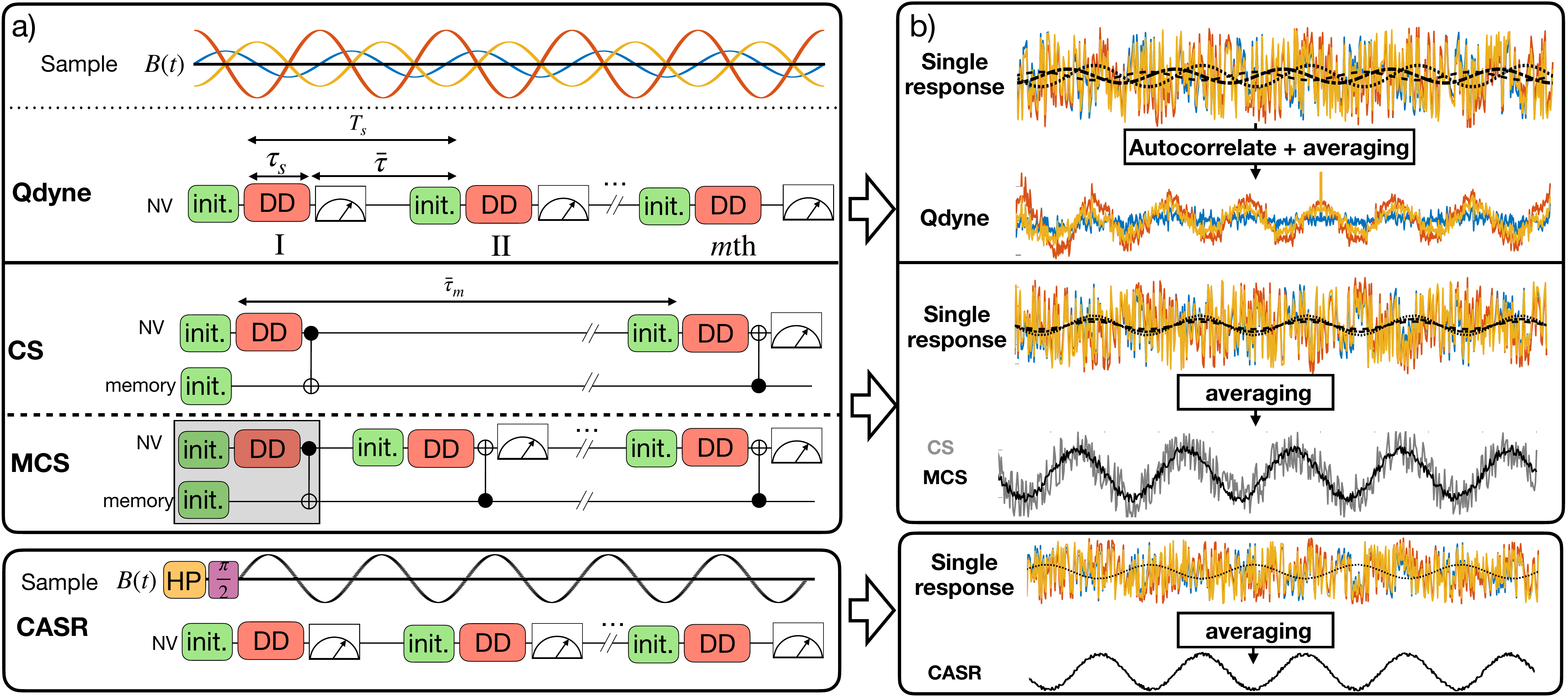}
    \caption{\textbf{a) Sensing schemes.} All DD blocks have duration $\tau_s$. Top: three realizations of the statistical-polarization signal, with random amplitude and phase. Qdyne repeats initialization--DD--readout blocks and autocorrelates the fluorescence record in post-processing. CS stores the phase from a first DD block in the memory qubit and correlates it with that of a second block after a waiting time $\bar{\tau}$. MCS correlates the stored reference (shaded) with $M$ successive blocks, avoiding long idle times. In CASR, hyperpolarization (HP) followed by an RF $\pi/2$ pulse creates a signal with the same phase in every realization, detected with synchronized DD blocks.
\textbf{b) Data processing.} Black lines: noiseless single-shot responses. Colored lines: measured noisy responses. In Qdyne, the random phase makes the raw traces average to zero, so each trace is autocorrelated before averaging. CS and MCS yield a phase-independent correlation, and CASR a fixed-phase signal, so their responses are averaged directly. MCS (black) is less noisy than CS (grey) because it accumulates more averages, see Sec.~\ref{sec:snr}.}
    \label{fig:fig1}
\end{figure*}
\section{Detection methods}\label{sec:2}
In this section, we describe the four methods considered in this work: Qdyne, CS, MCS and CASR.  Table~\ref{tab:methods} classifies them according to the \emph{what}, \emph{how} and \emph{when} choices introduced in Sec.~\ref{Intro}, and Fig.~\ref{fig:fig1}(a) illustrates the corresponding sequences. For each method, we derive the expectation value of the NV readout in the ideal, noiseless case, which shows how the signal parameters are encoded in the measured response, and we identify the reproducible, phase-independent quantity that is averaged over runs. Figure~\ref{fig:fig1}(b) illustrates where this quantity is obtained, i.e., the \emph{when} column of Table~\ref{tab:methods} As shown by the noisy responses, the measured signal is affected by readout noise. The effect of this noise on the averaged quantity, which determines the SNR, is analyzed in Sec.~\ref{sec:snr}.
\begin{table}[b]
\caption{Classification of the four methods according to \textit{what} contribution of the sample polarization they target, \textit{how} they acquire the time trace, and \textit{when} they remove the random phase of the statistical contribution. CASR requires no correlation.}
\label{tab:methods}
\small
\setlength{\tabcolsep}{5pt}
\begin{ruledtabular}
\begin{tabular}{cccc}
\shortstack{\textbf{Method}\\\phantom{(\textit{what})}} &
\shortstack{\textbf{Target}\\(\textit{what})} &
\shortstack{\textbf{Acquisition}\\(\textit{how})} &
\shortstack{\textbf{Correlation}\\(\textit{when})} \\
\hline
\textbf{Qdyne} & Statistical & Multipoint   & After readout \\
\textbf{CS}    & Statistical & Single-point & Before readout \\
\textbf{MCS}   & Statistical & Multipoint   & Before readout \\
\textbf{CASR}  & Mean        & Multipoint   & Not required \\
\end{tabular}
\end{ruledtabular}
\end{table}

We consider an ensemble of $K$ shallow NV centers located beneath a nuclear ensemble (the sample to be sensed). Each NV is sensitive only to the $N$ nuclear spins closest to it. These spins precess about the applied static magnetic field and generate an oscillating magnetic field at the position of the NV, which contains both a mean and a statistical contribution of the nuclear polarization. Which of the two is sensed and survives the averaging over runs is set by the protocol, i.e., by the \emph{what} choice introduced in Sec.~\ref{Intro}. For either contribution, the component of this field along the NV axis is
\begin{align}
    B(t) &= b_C \cos\big[(\omega_C+\delta)t+\epsilon_C \big],
    \label{nuclear_field}
\end{align}
where $\omega_C = 2\pi\nu_C$ is the nuclear Larmor frequency and $\delta\ll\omega_C$ is the chemical shift. For statistical detection, we take $\epsilon_C$ uniformly distributed and replace the random amplitude by the constant $b_C\approx 300$ nT~\cite{spohn2025quantum}, as depicted in Fig.~\ref{fig:fig1}(a) (top). For mean detection, both the amplitude and phase are the same accros different experimental runs, as depicted in Fig.~\ref{fig:fig1}(a) (bottom). The amplitude of the mean contribution is defined in terms of $b_C$ in Sec. \ref{subsec:casr}. Finally, we assume that the statistical field keeps its phase during each experimental trace.

The goal is to extract $\delta$ from the interaction between $B(t)$ and the NV center. To this end, a MW control implements a dynamical-decoupling (DD) sequence whose filter function is centered at the nuclear precession frequency $\omega_C$. The DD blocks are applied stroboscopically, with their starting times separated by $T_s$, where $\omega_C T_s = 2\pi q$ and $q$ is an integer (see Fig.~\ref{fig:fig1}). The carrier at $\omega_C$ then accumulates no net phase between any two blocks. Within the rotating-wave approximation ($\omega_C\gg\gamma_{\rm NV} b_C$), the interaction of an individual NV center with the nuclear field during every DD block is therefore described by the same effective Hamiltonian (see App.~\ref{app:A})
\begin{equation}
H_{\rm{eff}}(t)= g_{\rm{DD}} \cos(\delta t+\epsilon_{\rm C})\sigma_z, \qquad g_{\rm{DD}}= \frac{\gamma_{\rm{NV}}b_{\rm C}}{\pi},
\label{app_Heff1}
\end{equation}
where $\sigma_z$ is the Pauli matrix of the $\{|0\rangle,|1\rangle\}$ manifold of the NV, and $t$ is measured from the start of the experimental run. This Hamiltonian is the common starting point for all the sensing protocols.

\subsection{Statistical detection methods}
\textit{Quantum heterodyne (Qdyne)}~\cite{schmitt2017submillihertz,boss2017quantum}. The method consists in repeated DD sensing blocks to track the temporal evolution of the nuclear magnetic field, as shown in Fig.~\ref{fig:fig1}(a). Each measurement cycle consists of a DD block of duration $\tau_s$ followed by a dead time $\bar{\tau}$, which contains the NV readout and its reinitialization for the next cycle. The cycles are thus repeated with period $T_s=\tau_s+\bar{\tau}$, thus the $m$th DD block starts at $t_m=mT_s$. The full time trace is therefore acquired within a single experimental run (multipoint acquisition). After the $m$th cycle, the response of the $k$th NV in the $r$th run is (see App.~\ref{app:A})
\begin{equation}
S_m^{(rk)} \simeq A_{\rm Q} \cos\left( \delta t_m+\epsilon_{\rm C}^{(rk)} \right),
\label{Qdyne_signal}
\end{equation}
with $A_{\rm Q}\simeq 2g_{\rm{DD}}\tau_s$ in the weak-field limit ($g_{\rm DD}\tau_s\ll 1$) and for $\delta\tau_s\ll 1$. To remove th dependence on $\epsilon_C^{(rk)}$, the measured time trace of each run is auto-correlated,
$\displaystyle
C_\ell^{(rk)}~=~\frac{1}{M}\sum_{m=1}^{M-\ell} S_m^{(rk)} S_{m+\ell}^{(rk)},
$
where $M$ is the number of sensing blocks in the trace and $\ell\geq 1$ is the lag between blocks. In the absence of noise, averaging over different runs gives (see App.~\ref{app:qdyne_corr})
\begin{equation}
\overline{C_\ell} \simeq \frac{A_{\rm Q}^2}{2} \left(1-\frac{|\ell|}{M}\right)\cos(\delta \ell T_s).
\label{qdyne_autocorr}
\end{equation}
Eq.~\eqref{qdyne_autocorr} thus represents the effective phase-independent signal that can be compared with the other detection methods. 

\textit{Correlation Spectroscopy (CS)}~\cite{laraoui2013high}. In CS, two identical DD sensing blocks are separated by a variable correlation delay $\bar{\tau}_m$. Here we consider the memory-assisted version of CS~\cite{rosskopf2017quantum}, in which the phase acquired in the first block is stored during the delay in the populations of a nuclear quantum memory using a ${\rm C_eNOT_n}$ gate (where e=electron and n=nucleus), whose relaxation time is much longer than the $T_1$ of the NV. After the second DD block, the information is retrieved from the memory with a ${\rm C_nNOT_e}$ and the NV is read out. Unlike in Qdyne, the correlation between the two sensing blocks is thus encoded in the quantum state of the sensor before the readout, and no autocorrelation is needed in post-processing. Since each run yields the correlation at a single delay, the time trace is acquired point by point by sweeping $\bar{\tau}_m = \bar\tau_0 + m\Delta \bar\tau$ over successive runs (single-point acquisition). The corresponding NV response in the limit $\delta\tau_s\ll 1$ is (see App.~\ref{app:A})
\begin{equation}
S_m^{(rk)} \simeq A_{\rm{CS}} \left[ \cos(\delta\bar{\tau}_m)+\cos\left(2\epsilon_{\rm C,m}^{(rk)}+\delta\bar{\tau}_m\right)\right]; \ A_{\rm{CS}}=\frac{A_{\rm Q}^2}{2}.
\label{eq:cs_productsignal}
\end{equation}
The first term does not depend on the random phase and therefore survives averaging over runs. The second term vanishes on average. By construction of the method, however, its fluctuations do not contribute to the signal but add to the noise background, which reduces the SNR (see Sec.~\ref{sec:snr} and Appendix~\ref{app:snr_cs_mcs}).

\textit{Multipoint correlation spectroscopy (MCS)}~\cite{spohn2025quantum,maier2025efficient}. MCS combines the main idea of CS, namely correlating the signal with a stored reference before the readout, within the quantum state of the sensor, with the multipoint acquisition of Qdyne. After an initial reference block, shaded in Fig.~\ref{fig:fig1} a), the acquired phase is mapped onto the populations of the quantum memory by a ${\rm C_eNOT_n}$ gate. The NV electron spin is then reinitialized and used to acquire further phases in $M$ successive DD blocks. After each of them, the newly acquired phase is correlated with the stored reference by a ${\rm C_nNOT_e}$ gate, and the NV is read out. Assuming that the memory is not disturbed by the $M$ optical readouts~\cite{neumann2010single,spohn2025quantum}, the NV response is writen as,
\begin{equation}
S_m^{(rk)} \simeq \frac{A_{\rm Q}^2}{2} \left[ \cos(\delta t_m)+\cos\left(2\epsilon_{\rm C}^{(rk)}+\delta t_m\right)\right].
\label{eq:app_mcs_productsignal}
\end{equation}
As in CS, the first term survives averaging over runs, whereas the second one vanishes. The second term again vanishes on average and, in contrast to CS, by construction of the method its fluctuations do not add to the noise background, leaving the SNR unaffected (see Sec.~\ref{sec:snr} and Appendix~\ref{app:snr}).

\subsection{Mean detection methods}
\label{subsec:casr}
\textit{Coherently averaged synchronized readout (CASR)}~\cite{glenn2018high}. Unlike the previous schemes, which detect the statistical polarization of the sample, CASR detects its mean polarization $p$, which is typically enhanced beyond its thermal value, i.e., $P_{\rm th}=\hbar\gamma B/(2k_BT)$, by hyperpolarization~\cite{von2025coherent}. As shown in Fig.~\ref{fig:fig1}(a), the sequence is the same as in Qdyne: DD blocks of duration $\tau_s$ are repeated at intervals $T_s=\tau_s+\bar{\tau}$, and the NV is read out after each block. The NV response is (see App.~\ref{app:A})
\begin{equation}
S_m^{(rk)} \simeq A_{\rm{CASR}} \cos(\delta t_m+\epsilon_{\rm C}), \qquad A_{\rm{CASR}} = \frac{b_{\rm C}^{\rm mean}}{b_{\rm C}}\,A_{\rm Q},
\label{casr_signal}
\end{equation}
where the ratio $b_{\rm C}^{\rm mean}/b_{\rm C}=p\sqrt{N}$~\cite{herzog2014boundary}. In contrast with the previous methods, the initial phase $\epsilon_{\rm C}$ is fixed by the RF $\pi/2$ pulse that starts the nuclear precession, and is therefore the same in every run and for every sensor. No correlation is thus required: the signal itself survives averaging over runs and sensors. The price is that the mean polarization must be re-established between runs, both by longitudinal relaxation of the sample and by repeating the hyperpolarization step.
\section{Signal-to-noise ratio}
\label{sec:snr}
\begin{table}[t]
\caption{
Protocol-dependent quantities entering the general SNR expression, Eq.~\eqref{eq:general_snr}. The spectral peak $\mathcal{P}_{\alpha}$ refers to the Fourier amplitude for Qdyne, CS, and MCS, and to the Fourier power for CASR. The statistical factors $\chi_{\rm{\alpha}}$s are defined in App.~\ref{app:snr}, \ref{app:snr_qdyne} and \ref{app:snr_casr}. $L_{\rm eff}$ is the maximum autocorrelation lag retained in Qdyne , and $A_{\rm{CS}}=A_{\rm{MCS}}=A_Q^2/2$.}
\label{tab:snr_parameters}

\begin{ruledtabular}
\begin{tabular}{lcccc}
Protocol ($\alpha$) &
$\mathcal{P}_{\alpha}$ &
$\chi_{\alpha}$ &
$\mathcal{N}_{\alpha}$ &
$\beta_{\alpha}$ \\
\hline
Qdyne &
$\widetilde A_Q^2/4$ &
$\sqrt{1-\frac{2}{\pi}}$ &
$\sigma_{\rm{meas}}^2/\sqrt{L_{\rm eff}}$ &
$1/2$ \\
CS &
$\widetilde A_{\rm{CS}} \gamma_1^{(\rm CS)}/2$ &
$\frac{\sqrt{4-\pi}}{2}$ &
$\sqrt{\sigma_{\rm{meas}}^2+
A_{\rm{CS}}^2/2}$ &
$1/2$ \\
MCS &
$\widetilde A_{\rm{MCS}}\gamma_1^{(\rm MCS)}/2$ &
$\frac{\sqrt{4-\pi}}{2}$ &
$\sigma_{\rm{meas}}$ &
$1/2$ \\
CASR &
$\widetilde A_{\rm{CASR}}^2/4$ &
$1$ &
$\sigma_{\rm{meas}}^2$ &
$1$
\end{tabular}
\end{ruledtabular}
\end{table}

Within a total experimental time $T_{\rm tot}$, the number of experimental repetitions is $R^{(\alpha)}=\lfloor T_{\rm tot}/T^{(\alpha)}_{\rm trace} \rfloor$, where $T^{(\alpha)}_{\rm trace}$ is the time needed to acquire one full trace for different methods $\alpha$ . 
We begin by computing the time $T_{\rm trace}$ needed to acquire and reconstruct a full time trace of $M$ points with a sampling interval of $T_s$. For the multipoint methods ($\alpha \in$Qdyne, MCS and CASR), a single run yields the full trace, so that
\begin{equation}
T_{\rm trace}^{(\alpha)} = t_{\rm prep}^{(\alpha)} + (M-1) T_s, 
\label{eq:trace_duration}
\end{equation}
where the preparation time is
\begin{equation}
t_{\rm prep}^{(\alpha)} =
\begin{cases}
t_{\rm init}, & \alpha=\text{Qdyne}, \\
t_{\rm ref}, & \alpha=\text{MCS}, \\
t_{\rm hp}, & \alpha=\text{CASR}.
\end{cases}
\label{eq:preparation_time}
\end{equation}
Here, $t_{\rm init}$ is the laser initialization time of the NV, $t_{\rm ref}=t_{\rm init}+\tau_s+t_{\rm CeNOTn}$ is the duration of the MCS reference block (gray box in Fig.~\ref{fig:fig1}(a)), and $t_{\rm hp}$ is the time needed to hyperpolarize the sample. In CS, however, each run yields a single point in the trace and the total time needed for the trace is therefore written as,  
\begin{align}
T_{\rm trace}^{\rm CS}
&=M\left(t_{\rm init}+2\tau_s+ \bar\tau_0 + t_{\rm ro}\right)+\frac{M(M-1)}{2}\,\Delta\bar\tau,
\label{eq:trace_duration_cs}
\end{align}
where, $t_{\rm ro}$ is the final readout of duration for each $m$. We neglect the rest of the remaining minor overheads, such as the gate times within each sensing block, control-pulse durations, and other protocol-specific delays, assuming that their contributions to the total acquisition time are negligible.

Having established the acquisition-time requirements for each protocol, we now quantify their sensing performance through the signal-to-noise ratio (SNR). Defining the SNR as the ratio of the signal peak to the standard deviation of the corresponding spectral background, a unified expression applicable to all four methods is
\begin{equation}
{\rm SNR}_{\alpha} = \frac{\mathcal{P}_{\alpha}}{\chi_{\alpha}\mathcal{N}_{\alpha}}\left(\frac{M K T_{\rm tot }}{ T_{\rm trace}^{(\alpha)}}\right)^{\beta_{\alpha}}.
\label{eq:general_snr}
\end{equation}
Here, $\mathcal{P}_{\alpha}$ denotes the signal peak of the fourier spectra, $\mathcal{N}_{\alpha}$ is the effective noise scale, and $\chi_{\alpha}$ accounts for the statistical distribution of the noisy spectral background. The exponent $\beta_{\alpha}$ distinguishes the amplitude-based SNR ($\beta_{\alpha}=1/2$) from the power-based SNR ($\beta_{\alpha}=1$). All the detailed derivations regarding each method is given in App.~\ref{app:snr_qdyne}, \ref{app:snr_cs_mcs} and \ref{app:snr_casr}. To account for finite coherence and relaxation times, we
introduce the NV dephasing factor $\gamma_2=\exp[-\tau_s/T_2^{\mathrm{NV}}]$, where $T_2^{\mathrm{NV}}$ is the NV coherence time under dynamical decoupling. Accordingly, the effective sensing amplitudes are $\widetilde A_Q=\gamma_2 A_Q$, $\widetilde A_{\rm CASR}=\gamma_2 A_{\rm CASR}$, and $\widetilde A_{\rm CS}=\widetilde A_{\rm MCS} =\gamma_2^2 A_Q^2/2$.

For CS and MCS, we additionally account for the longitudinal relaxation of the nuclear memory through the factors
\begin{equation}
\gamma_1^{(\alpha)} =\frac{1}{M}\sum_{m=0}^{M-1} \exp\left[ -\frac{\,t_{\mathrm{mem},m}^{(\alpha)}} {T_{1,\mathrm{nuc}}} \right],
\label{Eq:T1}
\end{equation}
where $t_{\mathrm{mem},m}^{(\alpha)}$ is the storage time associated with the $m$th measurement. These relaxation effects are incorporated into the protocol-dependent quantities summarized in Table~\ref{tab:snr_parameters}.

Furthermore, We  also quantify the performance of each protocol through its spin sensitivity, $\eta_N^{(\alpha)}=N_{\min}^{(\alpha)} \sqrt{T_{\rm tot}}$, where $N_{\min}^{(\alpha)}$ is the minimum number of nuclear spins sensed by each NV required to achieve a constant SNR value of $C_r$. For independent nuclear spins, $A_Q^2=Na_1^2$, where $a_1$ denotes the single-spin contribution. Consequently, the signal contribution in Eq.~\eqref{eq:general_snr} scales as $\mathcal{P}_\alpha(N)=N^{2\beta_\alpha} \mathcal{P}_\alpha(1)$, where $\beta_\alpha=1/2$ for Qdyne, CS, and MCS, and $\beta_{\rm CASR}=1$. Solving for $N$ gives the sensitivity as,
\begin{equation}
\eta_{N}^{(\alpha)}
=
\left[
\frac{C_r\chi_\alpha\mathcal{N}_\alpha}
{\mathcal{P}_\alpha(1)}
\right]^{\frac{1}{2\beta_\alpha}}
\sqrt{\frac{T_{\rm trace}^{(\alpha)}}{KM}}.
\label{eq:general_sensitivity}
\end{equation}
Using Eqs.~\eqref{eq:general_snr} and~\eqref{eq:general_sensitivity}, we can systematically analyze how the differences among the four protocols arise from their signal amplitudes, spectral-noise statistics, and acquisition times, which in turn govern their spin sensitivities. 

\section{Results}\label{sec:4}
We compare the four protocols under common conditions. All simulations use 
an XY8 sequence of eight ideal $\pi$ pulses, a nuclear Larmor frequency $\nu_{\rm C}~=~167\,\rm{kHz}$, a chemical shift $\delta=100\,\rm{Hz}$, and a statistical field amplitude $b_{\rm C}^{\rm stat}=300\,\rm{nT}$, produced by $N_{\rm nuc}=6\times10^4$ nuclear spins. We set the sensing time to $\tau_s~=~T_2^{\rm NV}~=~23.95\,\mu\rm{s}$, i.e., four nuclear precession periods. All protocols share the sampling period $T_s=14/\nu_{\rm C}\simeq83.83\,\mu\rm{s}$ ($\Delta\tau=T_s$ in CS), so that for a given $M = 1193$ they reach the same resolution $\Delta\nu_{\rm res}=1/(MT_s)\approx 10.0$ Hz. This minimizes spectral leakage and allows a consistent evaluation of the spectral peak across all four protocols. In addition, for CASR we assume a nuclear polarization $p=2\times10^{-4}$, which gives $b_{\rm C}^{\rm mean}\approx15\,\rm{nT}$~\cite{von2025coherent}. 

In CASR, Qdyne, and MCS, consecutive sensing blocks are separated by a dead time $\bar{\tau}=T_s-\tau_s\simeq59.88\,\mu\rm{s}$, which contains the optical readout and the NV re-initialization and, in MCS, also a $\rm C_nNOT_e$ gate. Each MCS trace further starts with a reference-sensing block followed by a $\rm C_eNOT_n$ gate. The duration of such gates are $T_{\rm C_nNOT_e}~\approx~1\,\mu\rm{s}$ and $T_{\rm C_eNOT_n}~\approx~50\,\mu\rm{s}$~\cite{spohn2025quantum}. Each CASR trace includes a hyperpolarization time $t_{\rm hp}=0.3\,\rm{s}$. In CS the delays start at $\bar\tau_0\simeq101.8\,\mu\rm{s}$ and increase in steps of $\Delta\tau=T_s$. For a total acquisition time $T_{\rm tot}=3600\,\rm{s}$, the number of complete traces, $R_\alpha= T_{\rm tot}/T_{\rm trace}^{(\alpha)}$, is $R_{\rm CASR}~=~9000$, $R_{\rm QD}=36013$, $R_{\rm MCS}=35968$, and $R_{\rm CS}=60$. Unless stated otherwise, we use $K=10^4$ NV centers and a readout noise $\sigma_{\rm meas}~=~20$~\cite{von2025coherent,hopper2018spin}.

\begin{figure}
    \centering
    \includegraphics[width=\linewidth]{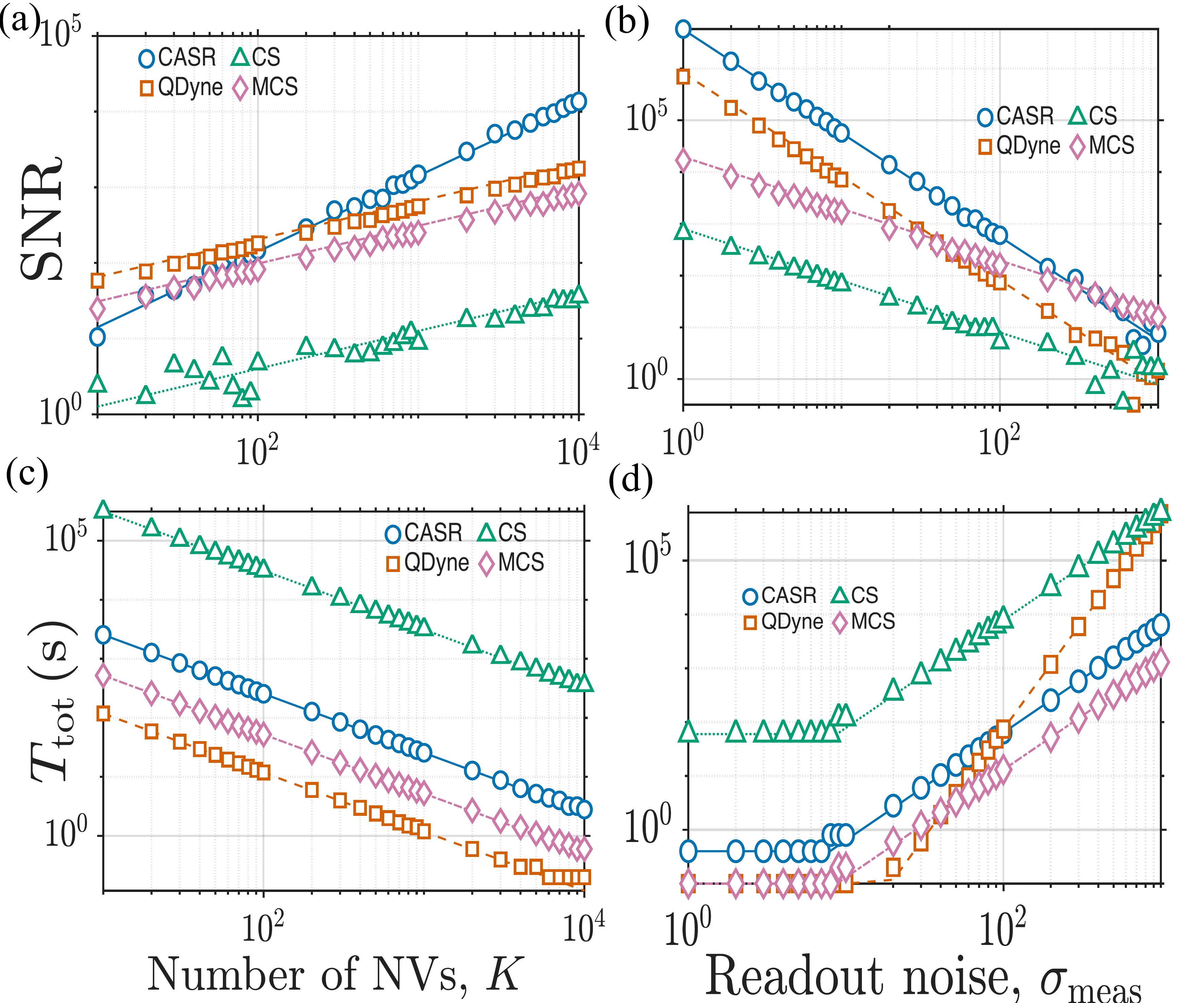}
    \caption{Comparison of the sensing performance of CASR (circles), Qdyne (squares), CS (triangles), and MCS (diamonds). (a) SNR versus the number of NV centers $K$ at $\sigma_{\rm meas}=20$. (b) SNR versus the readout-noise standard deviation $\sigma_{\rm meas}$ at $K=10^4$. Both panels use $T_{\rm tot}=3600\,\rm{s}$. (c) Total acquisition time $T_{\rm tot}$ required to reach $\mathrm{SNR}=10$ versus $K$ at $\sigma_{\rm meas}=20$. (d) Same as (c) versus $\sigma_{\rm meas}$ at $K=10^4$. In (c) and (d), $T_{\rm tot}=RT_{\rm trace}$, with the number of traces $R$ rounded up to the nearest integer. Symbols show numerical results and lines the analytical predictions of Eq.~\eqref{eq:general_snr}. For Qdyne, the maximum autocorrelation lag is $L=M-1$. The nuclear-memory relaxation time is $T_{1,\rm nuc}=0.1\,\rm{s}$.}
    \label{fig:fig2}
\end{figure}

Figure~\ref{fig:fig2}(a) shows that the SNR grows as $\sqrt{K}$ for Qdyne, CS, and MCS, but linearly in $K$ for CASR. This steeper scaling arises because the CASR SNR is defined on the Fourier power of a coherently accumulated signal, which grows linearly with the number of independent measurements, whereas the statistical protocols use the Fourier amplitude of a correlation, which grows as its square root (Sec.~\ref{sec:snr}). In terms of spin sensitivity, all four protocols scale as $K^{-1/2}$, (see Eq.~\eqref{eq:general_sensitivity}). CASR thus yields the largest SNR for $K\gtrsim200$. Among the statistical protocols, Qdyne achieves the highest SNR by averaging each trace over many autocorrelation lags, while MCS exceeds CS by a factor of $\approx20$, since repeatedly accessing the stored nuclear phase allows many more measurements within the same time budget, see the averaged CS and MCS responses in Fig.~\ref{fig:fig1}(b). Figure~\ref{fig:fig2}(b) shows that the SNR decreases as $\sigma_{\rm meas}^{-2}$ for CASR and Qdyne, where the readout noise enters quadratically (through the background of the Fourier power and through noise-noise products in the autocorrelation, respectively), but only as $\sigma_{\rm meas}^{-1}$ for CS and MCS, where the correlation is established in the memory before the readout. Consequently, MCS overtakes Qdyne for $\sigma_{\rm meas}\gtrsim40$ and CASR for $\sigma_{\rm meas}\gtrsim400$.

In Figs.~\ref{fig:fig2}(c) and (d) we show the time required to reach a target SNR of $10$. For each protocol, the minimum number of complete repetitions is determined independently, giving $T_{\rm tot}=RT_{\rm trace}$. Specifically, Fig.~\ref{fig:fig2}(c) shows that the required acquisition time decreases inversely with $K$ for all methods. The single-point trace construction of CS implies longer experimental times. In contrast, Qdyne, MCS, and CASR acquire the entire trace in a single run, substantially reducing the experimental overhead. Note also that CASR requires more time than Qdyne and MCS, even though it yields the largest SNR in Fig.~\ref{fig:fig2}(a) for $K\gtrsim200$. Since the CASR SNR grows linearly with $T_{\rm tot}$, rather than as $\sqrt{T_{\rm tot}}$, this advantage is lost when the target SNR is reached in much less than $3600\,\rm{s}$, fact which is also observable in the sensitivity analysis, see Eq.~\eqref{eq:general_sensitivity} and Fig.~\ref{fig:fig4}.

Figure~\ref{fig:fig2}(d) shows the acquisition time as a function of measurement noise. In the measurement-noise-dominated regime, the required time scales as $\sigma_{\rm meas}^4$ for Qdyne, but only as $\sigma_{\rm meas}^2$ for CS and MCS, where correlations are established before readout. Although the CASR SNR also depends quadratically on measurement noise, Eq.~\eqref{eq:general_snr}, its linear dependence on $T_{\rm tot}$ results in the same quadratic noise scaling of the required time as in CS and MCS. At large measurement noise, MCS requires substantially less acquisition time than the other methods (it is the fastest for $\sigma_{\rm meas}\gtrsim40$), making it the most efficient protocol in this regime. Conversely, at sufficiently low noise, a single complete trace already exceeds the target $\mathrm{SNR}=10$, so the acquisition time cannot be reduced further: it is set by the duration of that one trace ($\approx0.1\,\rm{s}$ for Qdyne and MCS, $0.4\,\rm{s}$ for CASR, and $60\,\rm{s}$ for CS) and no longer depends on the noise level. This gives rise to the plateaus observed at small $\sigma_{\rm meas}$. 
Importantly, all these features are accurately reproduced by our analytical expressions, which show near-perfect agreement with the numerical simulations across all four protocols and the parameter ranges considered. This agreement confirms the validity of the analytical framework in describing the SNR and $T_{\rm tot}$ under different experimental conditions.
\begin{figure}
    \centering
    \includegraphics[width=\linewidth]{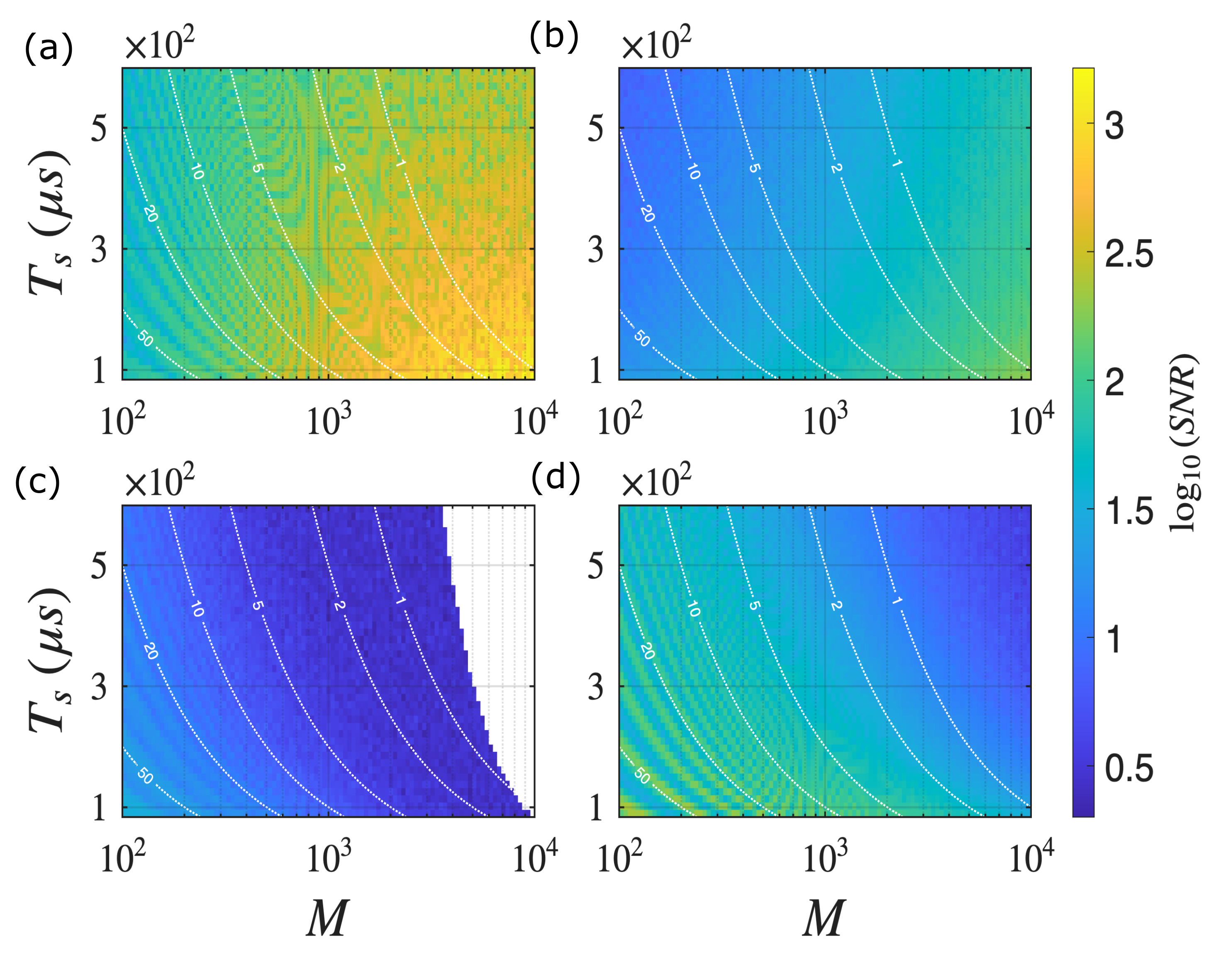}
    \caption{SNR as a function of $M$ and the sampling interval $T_s$ for (a) CASR, (b) Qdyne, (c) CS, and (d) MCS. The color scale represents $\log_{10}(\rm{SNR})$, with a common scale across all four panels. White dotted contours indicate constant nominal spectral resolution, $\Delta\nu_{\rm res}=1/(MT_s)$ (labels in Hz). The total acquisition time is fixed at $T_{\rm tot}=3600\,\rm{s}$, with $K=10^4$ NV centers and a Gaussian readout-noise standard deviation $\sigma_{\rm meas}=20$. Regions where a complete trace cannot be acquired within the available time budget are excluded. All other parameters are the same as in Fig.~\ref{fig:fig2}.}
    \label{fig:fig3}
\end{figure}

Fig.~\ref{fig:fig3} shows the SNR as a function of $M$ and $T_s$, with the white dotted contours marking constant frequency resolution, $\Delta\nu_{\rm res}=\frac{1}{M T_s}$. For all protocols, the SNR increases toward short $T_s$, since a shorter sampling period, and thus shorter time trace, leads to a higher number of experimental averages $R$. Along a contour, the trace duration $MT_s$, the number of traces $R$, and the memory factor $\gamma_1$ are all constant. The SNR of CASR and Qdyne thus grows toward large $M$ and short $T_s$, as each trace contains more samples. The SNR of MCS also increases along a contour, but more weakly, approximately as $\sqrt{M}$. In CS, the SNR is constant along each contour. At fixed $T_s$, increasing $M$ raises the SNR of Qdyne, which averages over more lags, and that of CASR until $MT_s\approx t_{\rm hp}$, where the gain saturates. It lowers the SNR of CS, for which fewer traces fit within $T_{\rm tot}$, and that of MCS once $MT_s$ exceeds $T_{1,\rm nuc}$. Large values of $M$ and $T_s$ are not accessible for CS within $T_{\rm tot}$ (white region). We stress that this analysis neglects the decoherence of the sample, i.e., the nuclear signal is assumed to keep its phase over the entire trace. In practice, its coherence time is limited by molecular diffusion, which randomizes the phase of the statistical polarization probed by the correlation-based protocols, or by dipolar dephasing in solid samples, which affects all four protocols. This bounds the achievable resolution and lowers the SNR whenever $MT_s=1/\Delta\nu_{\rm res}$ exceeds the coherence time. Figure~\ref{fig:fig3} is therefore not meant to identify the achievable resolution, but to show how the SNR changes with the targeted resolution, e.g., how much SNR can be gained by accepting a worse resolution.
\begin{figure}[b]
    \centering
    \includegraphics[width=\linewidth]{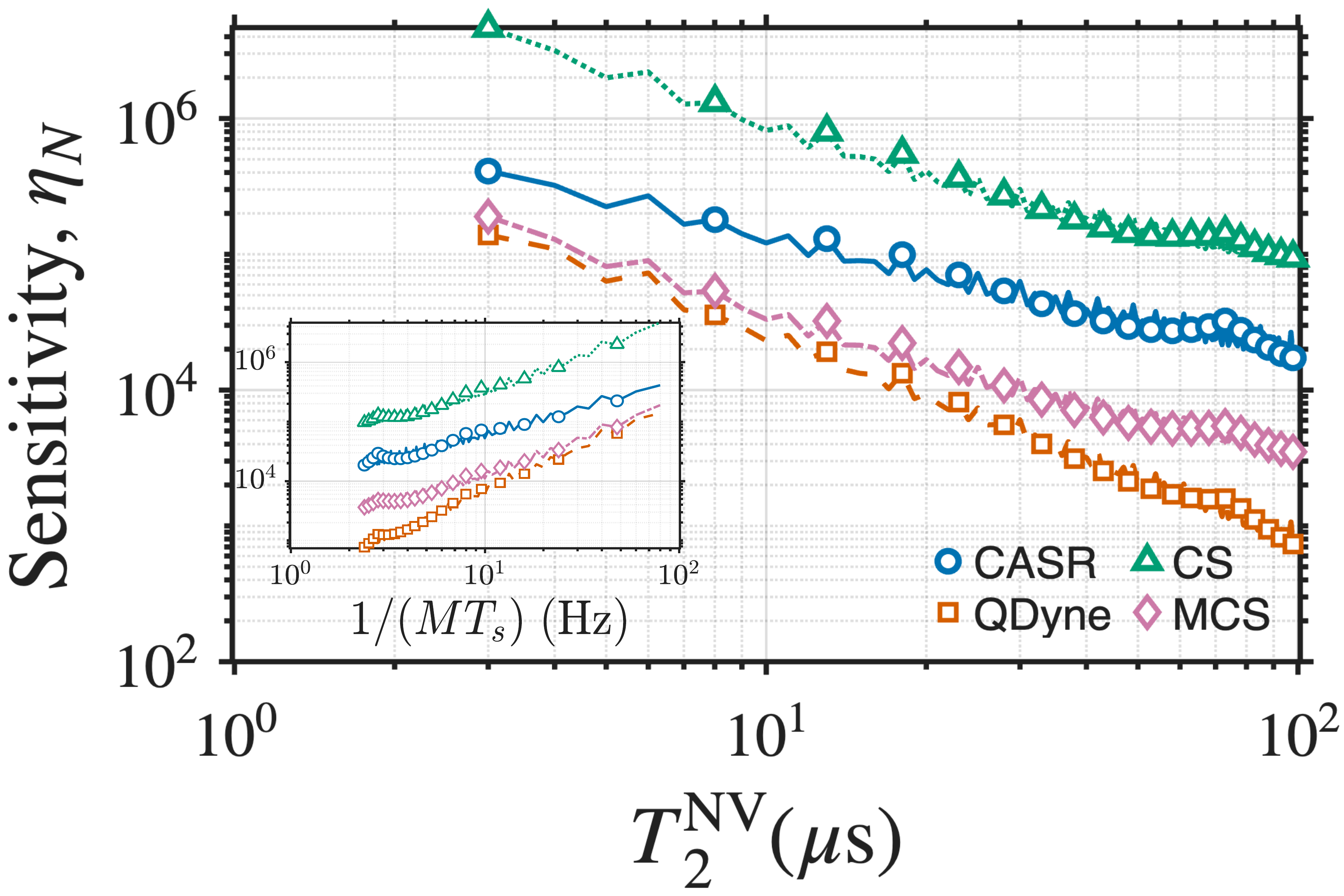}
    \caption{Spin sensitivity $\eta_N=N_{\min}\sqrt{T_{\rm tot}}$ as a function of the NV coherence time $T_2^{\rm NV}$ for CASR (circles), Qdyne (squares), CS (triangles), and MCS (diamonds). Here, $N_{\min}$ is the minimum number of nuclear spins required to reach a spectral SNR of $3$. The NV coherence time is varied from $3$ to $100\,\mu\rm{s}$, with $\tau_s=T_2^{\rm NV}$, $T_s=3.5\,\tau_s$, and $M=1193$. Inset: same data as a function of the nominal spectral resolution $\Delta\nu_{\rm res}=1/(MT_s)=1/(3.5\,MT_2^{\rm NV})$. Symbols show numerical results and lines the analytical predictions. All methods use $T_{\rm tot}=3600\,\rm{s}$, $K=10^4$ NV centers, $p=2\times10^{-4}$, and $\sigma_{\rm meas}=20$.}
    \label{fig:fig4}
\end{figure}

Figure~\ref{fig:fig4} presents the spin sensitivity $\eta_N=N_{\min}\sqrt{T_{\rm tot}}$, where $N_{\min}$ is the minimum number of nuclear spins needed to reach a spectral SNR of $3$, see Eq.~\eqref{eq:general_sensitivity}. The sensitivity improves with $T_2^{\rm NV}$ for all protocols, since a longer sensing time $\tau_s=T_2^{\rm NV}$ increases the accumulated phase. As $T_s$ is varied together with $\tau_s$ at fixed $M$, fewer samples fit within $T_{\rm tot}$ as $T_2^{\rm NV}$ grows, which yields $\eta_N\propto(T_2^{\rm NV})^{-3/2}$ for Qdyne, which is the most sensitive protocol, followed by MCS. The gap between the two, and that to CS, widens for $T_2^{\rm NV}\gtrsim24\,\mu\rm{s}$, where $MT_s$ exceeds $T_{1,\rm nuc}$. Note that CASR, which yields the largest SNR in Fig.~\ref{fig:fig2}, ranks only third. Figure~\ref{fig:fig2} refers to a sample of $N_{\rm nuc}=6\times10^4$ spins, whereas $N_{\min}\ll N_{\rm nuc}$, and the mean field detected by CASR, $b_{\rm C}^{\rm mean}\propto pN$, decreases faster than the statistical field, $b_{\rm C}^{\rm stat}\propto\sqrt{N}$, as $N$ is reduced. Since $M$ is fixed, the  resolution $\Delta\nu_{\rm res}$ also improves with $T_2^{\rm NV}$, and the inset of Fig.~\ref{fig:fig4} shows the same data as a function of $\Delta\nu_{\rm res}$. The better sensitivity at finer resolution is therefore inherited from the longer sensing time, rather than being an intrinsic benefit of a finer resolution, whose cost depends on the protocol (Fig.~\ref{fig:fig3}).

\section{Discussion}\label{sec:5}
\begin{figure}[t]
\includegraphics[width=\linewidth]{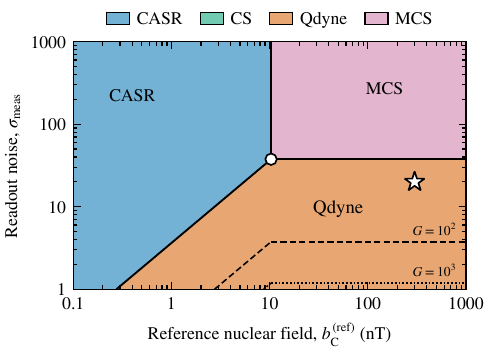}
\caption{Optimal protocol, i.e., the one with the lowest spin sensitivity $\eta_N$ [Eq.~\eqref{eq:general_sensitivity}] at SNR $=3$, as a function of the reference nuclear field $b_{\rm C}^{({\rm ref})}$ and the readout noise $\sigma_{\rm meas}$, for $p=2\times10^{-4}$. $b_{\rm C}^{({\rm ref})}$ is the statistical field produced by $N_{\rm ref}=6\times10^{4}$ nuclear spins and sets the single-spin signal $a_1=2\gamma_{\rm NV}\tau_s b_{\rm C}^{({\rm ref})}/(\pi\sqrt{N_{\rm ref}})$. Solid black lines mark where two protocols have the same $\eta_N$, assuming that each NV center is read out individually. Dashed (dotted) lines show the Qdyne frontiers when the NV centers are read out in groups of $G=10^2$ ($10^3$). For $G=10^4$ the Qdyne region lies entirely below $\sigma_{\rm meas}=1$. The circle marks the triple point and the star the working point of Figs.~\ref{fig:fig2}-\ref{fig:fig4} ($b_{\rm C}^{({\rm ref})}=300$\,nT, $\sigma_{\rm meas}=20$). CS is not optimal anywhere. The other parameters as in Fig.~\ref{fig:fig2}.}
\label{fig:map}
\end{figure}

As seen in the previous section, the behavior of each protocol can be traced back to the three choices introduced in Sec.~\ref{Intro} and Table~\ref{tab:methods}: \textit{what} contribution of the polarization it targets, \textit{how} it samples the signal, and \textit{when} it removes the random phase of the statistical contribution. As a result, no protocol is optimal under all experimental conditions, and the competition between them is captured by the ratios of their spin sensitivities, Eq.~\eqref{eq:general_sensitivity}.

Fig.~\ref{fig:map} summarizes which protocol reaches the lowest spin sensitivity as a function of the readout noise $\sigma_{\rm meas}$ and of the reference field $b_{\rm C}^{({\rm ref})}$, defined as the statistical field produced at the NV by $N_{\rm ref}=6\times10^{4}$ nuclear spins. Since $N_{\rm ref}$ is fixed, $b_{\rm C}^{({\rm ref})}$ measures the coupling between the NV and a single nuclear spin, i.e., the single-spin signal $a_1$. This coupling is set by the distance between the NV and the sample and by the targeted nuclear species, so it decreases for deeper NVs and for nuclear species with lower gyromagnetic ratios. It does not depend, instead, on the number of spins in the sample: since $\eta_N$ is defined through the minimum number of spins needed for detection, a change in the spin density leaves the map unchanged.

\textit{When the phase is removed: readout noise.} In Qdyne, the correlation is computed from the classical fluorescence record, i.e., after the noisy readout, Fig.~\ref{fig:fig1}(b) top panel. The background of the averaged autocorrelation is then dominated by noise--noise products, and $\eta_N^{\rm QD}\propto\sigma_{\rm meas}^2$. In CS and MCS, the correlation is performed before the noisy readout, Fig.~\ref{fig:fig1}(b) middle panels, and thus $\eta_N\propto\sigma_{\rm meas}$. Qdyne compensates for this by averaging over the $L_{\rm eff}$ effective lags of each trace. For the parameters of Sec.~\ref{sec:4}, $\eta_N^{\rm QD}/\eta_N^{\rm MCS}=1$ at $\sigma^{*}\approx27$, the horizontal boundary in Fig.~\ref{fig:map}. 

As anticipated by Figs.~\ref{fig:fig2}(b,d), Qdyne is more sensitive for $\sigma_{\rm meas}<\sigma^{*}$ and is surpassed by MCS for $\sigma_{\rm meas}>\sigma^{*}$. Assuming non-enhanced readout techniques~\cite{hopper2018spin}, conventional room-temperature readout of a single NV gives $\sigma_{\rm meas}\approx26$-$47$~\cite{von2025coherent,hopper2018spin}, so such experiments sit at or above $\sigma^{*}$, where MCS performs equally well or better than Qdyne. The advantage of Qdyne at low readout noise, however, relies on autocorrelating the fluorescence of each NV center individually~\cite{cambria2025scalable,cheng2025massively}. If the NV ensemble is read out in groups of $G$ centers whose fluorescence is detected collectively, the number of independent records drops from $K$ to $K/G$. This increases $\eta_N^{\rm QD}$ by an additional factor of $\sqrt{G}$ and lowers the Qdyne boundaries to $\sigma^{*}/\sqrt{G}$ (dashed lines in Fig.~\ref{fig:map}), while the other protocols are unaffected. At the working point, groups of $G\gtrsim4$ already favor MCS, and for a collective readout of the whole ensemble the SNR of Qdyne does not improve with $K$~\cite{spohn2025quantum}. 

\textit{How the signal is sampled.} As seen in Sec.~\ref{sec:4}, CS shares the linear noise scaling of MCS but acquires the trace point by point, so that its acquisition time grows quadratically with the number of points. The multipoint acquisition of the rest of the methods is thus far superior and MCS outperforms it everywhere in Fig.~\ref{fig:map}, which shows no CS region. Within our model, CS remains a reasonable choice only when a coarse resolution (small $M$) suffices, or when the memory does not survive the $M$ optical readouts assumed in MCS~\cite{neumann2010single,spohn2025quantum}, since in CS it only has to survive a single waiting time $\bar\tau_m$.

\textit{What is detected: polarization.} CASR detects the mean polarization, whose signal grows as $pN$ instead of the $N^{1/2}$ of the statistical field. As a result, $\eta_N^{\rm CASR}\propto(p\,a_1)^{-1}$, whereas the statistical protocols scale as $a_1^{-2}$, so the two classes respond differently to the single-spin coupling. A deeper NV needs more spins to reach the detection threshold, since each of them couples more weakly, but it also collects more spins within its detection volume. For CASR, both effects cancel for a uniformly polarized sample, whereas for the statistical protocols the number of required spins grows faster than the number of available ones. Weak couplings therefore favor CASR, whose region in Fig.~\ref{fig:map} lies at small $b_{\rm C}^{({\rm ref})}$, bounded by a vertical boundary with MCS and an oblique one with Qdyne, along which the readout noise also matters. At the working point of Figs.~\ref{fig:fig2}-\ref{fig:fig4}, CASR requires higher hyperpolarization enhancements to outperform the other methods. Such thing may be within reach of parahydrogen-induced polarization (PHIP) techniques, which have already been combined with NV-based NMR~\cite{arunkumar2021micron}.

\textit{Limitations.} Our analysis assumes Gaussian, uncorrelated readout noise, ideal pulses and gates, the weak-field regime, a statistical field whose phase is constant during each trace, and a memory unaffected by optical readouts.  More importantly for liquids, molecular diffusion through the detection volume randomizes the phase of the statistical field, which limits the useful trace length of the statistical protocols and hence their spectral resolution. The mean polarization detected by CASR is not dephased by diffusion, which is why CASR reaches higher resolutions~\cite{glenn2018high,von2025coherent} and is the only protocol considered here that can go beyond this diffusion limit. In solid samples, dipolar couplings between the nuclear spins dephase both the mean and the statistical polarization, and therefore limit the resolution of all four protocols.
\section{conclusions}\label{sec:conc}
To summarize, we have presented a unified theoretical and numerical framework for evaluating four NV-center-based nanoscale NMR detection protocols namely CASR, Qdyne, CS and MCS. By incorporating measurement noise, acquisition overheads, NV coherence, and nuclear-spin relaxation, we have derived analytical expressions for the SNR and spin sensitivity of each protocol. Our results demonstrate that the relative performance of these methods depends strongly on the experimental parameters and the available measurement resources. In particular, CASR benefits from coherent signal accumulation in the presence of finite nuclear polarization, whereas Qdyne exploits temporal correlations to resolve statistically polarized nuclear-spin signals. Among the correlation-based methods, Qdyne achieves best SNR and sensitivity by using classical correlation, while MCS is most resilient against the readout noise. We further show that the different dependencies on measurement noise and detectable statistical field lead to distinct operating regimes in which individual protocols become advantageous. These findings highlight the importance of optimizing the sensing protocol according to the available experimental parameters like nuclear polarization, readout noise, decoherences, and desired spectral resolution, . Overall, our study provides a quantitative framework for selecting and optimizing different nanoscale detection strategies in NV-center-based spectroscopy.

\begin{acknowledgments}
The authors thank Prof. Dominik Bucher for fruitful discussions. This work was supported by the European Union's Horizon Europe research and innovation programme under Grant Agreement No. 101135742 (QUENCH). J. C. acknowledges support from the Agencia Estatal de Investigaci'n through the project Modelizado, Optimizaci'n, y Esquemas de Magnetometría en Centros de Color (PID2024-161371NB-C21) and from the Basque Government under Grant No. IT1887-26. P. A. B. acknowledges financial support from UPV/EHU through the doctoral fellowship PIF 23/275.
\end{acknowledgments}

\bibliography{NV_sensing_references}

\onecolumngrid
\newpage
\begin{appendices}
\appendix

\section{Derivation of the effective Hamiltonian and single-block sensing response}
\label{app:A}

We consider an NV center interacting with the classical nuclear magnetic field defined in Eq.~\eqref{nuclear_field}. The Hamiltonian of the NV center under an applied static magnetic field and microwave driving can be written as \cite{doherty2013nitrogen, degen2017quantum, laraoui2013high}
\begin{equation}
H_{\rm Lab}(t) = \Delta S_z^2+\Omega_0 S_z
+\gamma_{\rm NV}B(t)S_z
+\sqrt{2}\Omega_1 f(t)
\cos(\omega_1 t+\varphi)S_x,
\label{NV_H}
\end{equation}
where $\Delta$ is the zero-field splitting, $\Omega_0=\gamma_{\rm NV}B_0$ is the Zeeman shift due to the static magnetic field $B_0$ along the NV axis, and $\Omega_1$ is the microwave Rabi frequency. The function $f(t)\in\{0,1\}$ is nonzero only during microwave control pulses. The term $\mathcal{B}_n(t)$ is the classical time-dependent magnetic field generated by the nuclear bath. We choose the manifold of $\ket{m_s=0}$ and $\ket{m_s=+1}$ such that $\sigma_z=\ket{+1}\bra{+1}-\ket{0}\bra{0}$. Transforming to the rotating frame of $\omega_1$ and applying the rotating-wave approximation under the resonance condition $\Omega_0+\Delta=\omega_1$, we obtain ($\Omega_C=\gamma_{\rm NV}b_C$)
\begin{equation}
H^{\rm RWA}(t) =
\frac{\Omega_C}{2}
\cos\left[(\omega_C+\delta)t+\epsilon_C\right]\sigma_z
+\frac{\Omega_1}{2}f(t)
\left[\cos\varphi\sigma_x
+\sin\varphi\sigma_y\right].
\label{RWA_twolevel_hamiltonian}
\end{equation}
In the pulsed dynamical-decoupling scheme, the microwave control is implemented through a sequence of ideal, instantaneous $\pi$ pulses about axes in the transverse plane. Each pulse reverses the sign of the longitudinal coupling, since $R_\alpha^\dagger(\pi)\sigma_z R_\alpha(\pi)=-\sigma_z$, for $\alpha=x,y$ depending on the choice of $\varphi$. Consequently, transforming to the toggling frame defined by the control sequence yields the Hamiltonian \cite{Cywinski2008PRB,deLange2011PRL,hernandez-gomez-2021}
\begin{equation}
H_P(t) =
\frac{\Omega_C}{2}y(t)
\cos\left[(\omega_C+\delta)t+\epsilon_C\right]\sigma_z,
\label{pulsed_hamiltonian}
\end{equation}
where $y(t)=(-1)^{n(t)}$ is the modulation function, with $n(t)$ denoting the number of $\pi$ pulses applied up to time $t$. For equally spaced pulses centered at $t_j=(j-\frac{1}{2})\tau_\pi$, we impose the resonance condition $\tau_\pi=\pi/\omega_C$. For a narrowband signal satisfying $|\delta|\ll\omega_C$, the modulation function is approximated by the first harmonic of its Fourier expansion, $y(t)=\operatorname{sgn}[\cos(\omega_C t)]\approx (4/\pi)\cos(\omega_C t)$. This gives the effective sensing Hamiltonian
\begin{equation}
H_{\rm eff}(t)=
g_{\rm DD}\cos(\delta t+\epsilon_C)\sigma_z,
\qquad
g_{\rm DD}=\frac{\gamma_{\rm NV}b_C}{\pi}.
\label{app_Heff2}
\end{equation}
This effective Hamiltonian provides a unified description of pulsed dynamical-decoupling sensing under ideal resonant conditions. We now use it to derive the accumulated phase and measurement response of an individual sensing block.

We consider the XY8 dynamical-decoupling sequence as the elementary sensing block employed in the different spectroscopy protocols \cite{Souza2011}. The sequence consists of eight $\pi$ pulses applied about the transverse axes in the order $(X,Y,X,Y,Y,X,Y,X)$. For ideal instantaneous pulses, the longitudinal modulation is identical to that of an equally spaced $\pi$-pulse sequence, while the alternating rotation axes provide robustness against pulse imperfections. We consider $N_8$ repetitions of XY8, corresponding to $N=8N_8$ pulses and a total sensing duration $\tau_s=N\tau_\pi$.

To calculate the accumulated sensing phase, we start from the toggling-frame Hamiltonian in Eq.~\eqref{pulsed_hamiltonian}. For the $m$th sensing block beginning at $t_m$, we introduce the local time $u=t-t_m$. The corresponding evolution operator is
\begin{equation}
U_m =
\exp\left[-i\int_0^{\tau_s}H_P(u)\,du\right]
=\exp\left(-i\frac{\Phi_m}{2}\sigma_z\right),
\label{eq:B1}
\end{equation}
where $\Phi_m$ is the accumulated relative phase between the two NV basis states. For equally spaced instantaneous $\pi$ pulses, the pulse positions within the sensing block are $u_j=(j-\frac{1}{2})\tau_\pi$, with $j=1,\ldots,N$. Defining $u_0=0$ and $u_{N+1}=\tau_s$, the modulation function takes the constant value $(-1)^j$ in each interval $[u_j,u_{j+1})$. Consequently, the accumulated phase can be evaluated by integrating over the individual pulse intervals,
\begin{align}
\Phi_m &=
\Omega_C\sum_{j=0}^{N}(-1)^j
\int_{u_j}^{u_{j+1}}
\cos(\omega u+\epsilon_C^m)\,du
\nonumber\\
&=
\frac{\Omega_C}{\omega}
\sum_{j=0}^{N}(-1)^j
\big[
\sin(\omega u_{j+1}+\epsilon_C^m)
-\sin(\omega u_j+\epsilon_C^m)
\big],
\label{eq:B2}
\end{align}
where $\omega=\omega_C+\delta$ and $\epsilon_C^m=\delta t_m+\epsilon_C$, assuming synchronization of the sensing blocks with the carrier frequency. Evaluating the sum over the pulse intervals yields
\begin{align}
\Phi_m &=
\frac{2\Omega_C}{\omega}
\left[
1-\sec\left(\frac{\omega\tau_\pi}{2}\right)
\right]
\sin\left(\frac{\omega\tau_s}{2}\right)
\cos\left(
\epsilon_C^m+\frac{\omega\tau_s}{2}
\right).
\label{eq:B3}
\end{align}
The apparent singularity at the resonance condition $\omega\tau_\pi=\pi$ is removable. Using $\omega_C\tau_\pi=\pi$ and considering the narrowband limit $|\delta|\ll\omega_C$, Eq.~\eqref{eq:B3} reduces, to leading order, to
\begin{equation}
\Phi_m \simeq
\frac{2\Omega_C\tau_s}{\pi}
\operatorname{sinc}\left(\frac{\delta\tau_s}{2}\right)
\cos\left(
\epsilon_C^m+\frac{\delta\tau_s}{2}
\right),
\label{eq:B4}
\end{equation}
where $\operatorname{sinc}(x)=\sin(x)/x$. The same result can also be obtained directly from the effective Hamiltonian in Eq.~\eqref{app_Heff1}. For ideal initialization and readout in the appropriate measurement quadrature, the resulting signal is \cite{degen2017quantum}
\begin{equation}
S_m=\langle\sigma_z\rangle_m=\sin\Phi_m.
\label{eq:B6}
\end{equation}
In the weak-signal regime, $|\Phi_m|\ll1$, the response becomes linear, $S_m\simeq\Phi_m$. This result describes the response of an individual XY8 sensing block, independent of the temporal sampling or correlation scheme. In the following appendices, we use this elementary response to construct the protocol-specific signals for the different spectroscopy methods.

\section{Autocorrelation of the Qdyne signal}
\label{app:qdyne_corr}
Here we derive the autocorrelation of the Qdyne signal. As discussed in the main text, each sensing block has duration $\tau_s$ and is repeated at intervals $T_s=\tau_s+\bar{\tau}$, with $t_m=mT_s$ and $m=0,\ldots,M-1$. We assume stroboscopic sampling, $\omega_{\rm{C}}T_s=2\pi q$, where $q$ is an integer, such that the sampled signal oscillates at $\delta$. In the weak-field regime, the signal acquired by the $k$th NV center during the $r$th experimental realization is
\begin{equation}
S_m^{(rk)} \simeq A_Q \cos\left( \delta mT_s+\epsilon_{\rm C}^{(rk)} +\frac{\delta\tau_s}{2} \right),
\label{eq:app_qdyne_signal}
\end{equation}
For a discrete time series containing $M$ measurements, we define the biased autocorrelation estimator at a nonnegative integer lag $\ell$ as \cite{Bendat2010},
\begin{equation}
\widehat C_{\ell}^{(rk)} = \frac{1}{M} \sum_{m=0}^{M-\ell-1} S_m^{(rk)}S_{m+\ell}^{(rk)},
\label{eq:app_qdyne_xcorr}
\end{equation}
where $\ell=0,1,\ldots,L$, with $L<M$ denoting the maximum retained lag. The physical correlation delay is $\tau_\ell=\ell T_s$. Here we chose normalization by $M$, rather than $M-\ell$, corresponding to the biased autocorrelation. Substituting Eq.~\eqref{eq:app_qdyne_signal} into
Eq.~\eqref{eq:app_qdyne_xcorr},
\begin{equation}
\widehat C_{\ell}^{(rk)}= \frac{A_Q^2}{2M} \sum_{m=0}^{M-\ell-1} \left[ \cos(\delta\ell T_s) + \cos\left( 2\delta mT_s+\delta\ell T_s +2\epsilon_{\rm C}^{(rk)} +\delta\tau_s\right) \right].
\label{eq:app_qdyne_product}
\end{equation}
The first term is independent of the initial nuclear phase gives the correlated signal, whereas the second explicitly depends on $\epsilon_{\rm C}^{(rk)}$. Averaging over the uniformly distributed initial phase yields
\begin{equation}
C_\ell \simeq \frac{A_Q^2}{2}\left(1-\frac{|\ell|}{M}\right)\cos(\delta\ell T_s); \qquad 
\Braket{ \cos\left(2\delta mT_s+\delta\ell T_s+2\epsilon_{\rm C}^{(rk)}+\delta\tau_s\right)}_{\epsilon_{\rm C}}=0.
\label{eq:app_qdyne_phase_avg}
\end{equation}
 where we have used the symmetry $C_{-\ell}=C_\ell$ for a real-valued signal. The prefactor $1-|\ell|/M$ originates from the finite number of sample pairs and the biased normalization, rather than from a physical decay of the nuclear correlations.

\section{Measurement-noise model}
\label{noise_model}
To establish a common framework for evaluating the sensitivity of the different sensing protocols, we include statistical fluctuations associated with the measurement of the NV electronic spin. For the $r$th experimental realization and the $k$th NV center, the measured signal at the $m$th sampling point is written as
\begin{equation}
Y^{(rk)}_m=S_m^{(rk)}+\xi_{rk,m},
\label{eq:measured_signal}
\end{equation}
where $S_m^{(rk)}$ denotes the ideal normalized NV response and $\xi_{rk,m}$ represents the measurement noise \cite{hopper2018spin}. We model these fluctuations as independent, zero-mean Gaussian random variables with variance $\sigma_{\rm{meas}}^2$, such that
\begin{equation}
\begin{aligned}
\langle\xi_{rk,m}\rangle &= 0,\\
\langle\xi_{rk,m}\xi_{r'k',m'}\rangle
&=\sigma_{\rm{meas}}^2
\delta_{rr'}\delta_{kk'}\delta_{mm'}.
\end{aligned}
\label{eq:noise_statistics}
\end{equation}
Here, $\sigma_{\rm{meas}}$ characterizes the uncertainty of an individual normalized readout, including the effective fluctuations associated with spin-state detection. The Gaussian model provides an approximation to the measurement statistics and assumes that the noise is uncorrelated between successive measurements, NV centers, and independent experimental realizations.

When the measured responses can be averaged coherently over $R$ realizations and $K$ NV centers, the resulting measurement-noise variance is reduced to
\begin{equation}
\sigma_{\rm{avg}}^2
=\frac{\sigma_{\rm{meas}}^2}{RK}.
\label{eq:noise_averaged}
\end{equation}
For protocols involving random nuclear phases, however, the appropriate signal-processing operation must be performed before averaging over independent realizations. In particular, the nuclear-phase fluctuations are incorporated in the ideal response $S_m^{(rk)}$ and are distinguished from the independent measurement noise $\xi_{rk,m}$. This distinction is essential when evaluating the spectral signal and its associated noise statistics.

\section{General recipe for evaluating the spectral
signal-to-noise ratio}
\label{app:snr}
Here, we establish a common framework for evaluating the spectral signal-to-noise ratio (SNR) using the Fourier amplitude spectrum. For the $r$th experimental realization and the $k$th NV center, the measured signal at the $m$th sampling point is,
\begin{equation}
Y_m^{(rk)}=S_m^{(rk)}+\xi_m^{(rk)},
\label{eq:snr_measured}
\end{equation}
where $S_m^{(rk)}$ denotes the ideal NV response and $\xi_m^{(rk)}$ represents independent Gaussian measurement noise with zero mean and variance $\sigma_{\rm{meas}}^2$, as described in App.~\ref{noise_model}. The measured signal might be subjected to the protocol-dependent processing operation $\mathcal{T}\left[\{Y_m^{(rk)}\}\right]_n$. For example $\mathcal{T}$ denotes classical autocorrelation for Qdyne, while CS and MCS processes $Y_m^{(rk)}$ directly. The processed signals are subsequently averaged over $R$ independent realizations and $K$ NV centers,
\begin{equation}
Y_n=\frac{1}{RK} \sum_{r=1}^{R}\sum_{k=1}^{K}Y_n^{(rk)} =\langle Y_n\rangle+\zeta_n,
\label{eq:snr_average}
\end{equation}
where $\zeta_n$ represents the fluctuations around the ensemble-averaged response, including residual nuclear-phase fluctuations and measurement noise. For protocols involving random nuclear phases, the correlation processing must precede averaging to preserve the desired signal. For a uniformly sampled  sequence containing $N_s$ points, the discrete Fourier transform is defined as 
\begin{equation}
\tilde{Y}_j=\frac{1}{N_s}\sum_{n=0}^{N_s-1}Y_n e^{-i2\pi jn/N_s},
\label{eq:snr_fft}
\end{equation}
with frequencies $\omega_j=2\pi j/(N_s\Delta t)$, where $\Delta t$ is the sampling interval. For CS, MCS, and CASR, the Fourier transform is evaluated directly from the $M$ uniformly sampled measurement points, such that $N_s=M$ and $n=m$, with an appropriate reindexing when necessary. The sampling interval is $\Delta\bar{\tau}$ for CS and $T_s$ for MCS and CASR. In contrast, Qdyne employs the symmetric autocorrelation sequence over $-L\leq\ell\leq L$, giving $N_s=2L+1$, with sampling interval $T_s$. The Fourier amplitude is $\mathcal{A}_j=|\widetilde{Z}_j|$. We define the spectral SNR as
\begin{equation}
\rm{SNR}_{\rm{FFT}} =\frac{\mathcal{A}_{\rm{sig}}} {\sigma_{\mathcal{A}}},
\label{eq:snr_general}
\end{equation}
where $\mathcal{A}_{\rm{sig}}$ denotes the signal contribution at the $j^*$th bin containing $\delta$, and $\sigma_{\mathcal{A}}$ is the standard deviation of the Fourier amplitude noise floor. Numerically, the latter is obtained from the spectral background, excluding the signal peak and other deterministic contributions. To evaluate the noise floor analytically, we consider independent Gaussian fluctuations $\zeta_n\sim\mathcal{N}(0,\sigma_Z^2)$. The corresponding Fourier coefficients are,
\begin{equation}
\widetilde{\zeta}_j= \frac{1}{N_s}\sum_{n=0}^{N_s-1} \zeta_n e^{-i2\pi jn/N_s} =A_j+i B_j.
\label{eq:snr_noise_fft}
\end{equation}
For nonzero, non-Nyquist Fourier bins, the orthogonality of the trigonometric functions gives $\braket {A_j^2} =\braket {B_j^2} =\frac{\sigma_Y^2}{2N_s}\equiv v$ and $\braket {A_jB_j} = 0$. Thus, $A_j$ and $B_j$ are independent, zero-mean Gaussian random variables with joint distribution
\begin{equation}
p(A_j,B_j)=\frac{1}{2\pi v}\exp\left[-\frac{A_j^2+B_j^2}{2v}\right].
\label{eq:snr_joint_pdf}
\end{equation}
Consequently, the Fourier noise amplitude $\mathcal{A}_j=\sqrt{A_j^2+B_j^2}$ follows a Rayleigh distribution\cite{Bendat2010},
\begin{equation}
p(\mathcal{A}_j) =\frac{\mathcal{A}_j}{v} \exp\left[-\frac{\mathcal{A}_j^2}{2v}\right], \qquad \mathcal{A}_j\geq 0.
\label{eq:snr_rayleigh}
\end{equation}
Its mean and variance are
\begin{equation}
\begin{split}
\langle\mathcal{A}_j\rangle
&=\sqrt{\frac{\pi v}{2}},\\
\rm{Var}(\mathcal{A}_j)
&=\frac{4-\pi}{2}v.
\end{split}
\label{eq:snr_rayleigh_moments}
\end{equation}
Therefore, the standard deviation of the Fourier amplitude noise floor is
\begin{equation}
\sigma_{\mathcal{A}} =\frac{\sigma_Y\sqrt{4-\pi}}{2\sqrt{N_s}}.
\label{eq:snr_amplitude_std}
\end{equation}

Finally, consider an oscillatory signal $Y_n^{(0)}=A\cos(\delta n\Delta t+\phi)$, where $\delta$ coincides with a nonzero, non-Nyquist Fourier bin. Its positive-frequency Fourier amplitude is $A/2$. Using Eq.~\eqref{eq:snr_amplitude_std}, we obtain
\begin{equation}
{\rm SNR}_{\rm{FFT}} =\frac{A/2}{\sigma_{\mathcal{A}}} =\frac{A\sqrt{N_s}} {\sigma_Y\sqrt{4-\pi}}.
\label{eq:snr_final}
\end{equation}
This expression applies to independent Gaussian fluctuations in the processed signal. In general, correlation processing and random nuclear phases can introduce non-Gaussian fluctuations and temporal correlations. The corresponding spectral noise floor must therefore be evaluated from the protocol-specific measurement statistics.

\section{SNR for Qdyne using Auto-correlated signal}
\label{app:snr_qdyne}
Here, we derive the spectral SNR of Qdyne using the biased autocorrelation introduced in App.~\ref{app:qdyne_corr} and the general SNR definition in App.~\ref{app:snr} \cite{schmitt2017submillihertz, boss2017quantum}. For the $r$th realization and $k$th NV center, the measured signal is,
\begin{equation}
Y_m^{(rk)} =A_Q\cos(\delta mT_s+\phi^{(rk)}) +\xi_m^{(rk)},
\label{eq:qsnr_signal}
\end{equation}
Note that we need to include the measurement noise before the autocorrelating the signal. Here $\phi^{(rk)}=\epsilon_C^{(rk)}+\delta\tau_s/2$ and $\xi_m^{(rk)}$ denotes the independent Gaussian measurement noise defined in App.~\ref{noise_model}. For a trace containing $M$ measurements, the biased autocorrelation at lag $\ell\geq 0$ is,
\begin{equation} 
\widehat C_\ell^{(rk)} =\frac{1}{M}\sum_{m=0}^{M-\ell-1} Y_m^{(rk)}Y_{m+\ell}^{(rk)}.
\label{eq:qsnr_corr}
\end{equation}
To calculate the spectral noise, we can expand the autocorrelation into signal-signal, signal-noise, and noise-noise  contributions i.e., $C_\ell^{(rk)} =C_{\ell,S}^{(rk)} +C_{\ell,S\xi}^{(rk)} +C_{\ell,\xi\xi}^{(rk)}$. In the measurement-noise-dominated regime $A_Q^2\ll\sigma_{\rm{meas}}^2$, the last term dominates,
\begin{equation}
C_{\ell,\xi\xi}^{(rk)} =\frac{1}{M}\sum_{m=0}^{M-\ell-1} \xi_m^{(rk)}\xi_{m+\ell}^{(rk)}  -\sigma_{\rm{meas}}^2\delta_{\ell0}.
\label{eq:qsnr_noise}
\end{equation}
The last term represents the measurements-noise contribution at zero lag and subtracted from the mean noise contribution. For non-negative lags, it is straight forward to obtain, 
\begin{equation}
\braket{ C_{\ell,\xi\xi}^{(rk)} C_{\ell',\xi\xi}^{(rk)} } = \frac{\sigma_{\rm{meas}}^4}{M^2} (M-\ell) (1+\delta_{\ell0}) \delta_{\ell\ell'}.
\label{eq:qsnr_corr_cov}
\end{equation}
This is similar to the variance in Eq.~\eqref{eq:noise_averaged}, specifically for Qdyne after averaging over $R$ independent realizations and $K$ individually resolved NV centers. Fourier transform of the symmetric correlation sequence, $C_{-\ell}=C_\ell$, using $N_s=2L+1$ points gives,
\begin{equation}
\widetilde C_j =\frac{1}{N_s} \left[C_0+2\sum_{\ell=1}^{L} C_\ell\cos\left(\frac{2\pi j\ell}{N_s}\right)\right].
\label{eq:qsnr_fft}
\end{equation}
For a signal frequency matching a nonzero , non-Nyquist Fourier bin, the spectral peak is 
\begin{equation}
\mathcal A_{\rm{sig}} = \left|\widetilde{\langle C\rangle}_{j_\delta}\right| \simeq\frac{A_{\rm QD}}{2} =\frac{A_Q^2}{4},
\label{eq:qsnr_peak}
\end{equation}
The Fourier noise variance can be expressed as
\begin{equation}
V_j\simeq \frac{\sigma_{\rm{meas}}^4} {RK\,ML_{\rm eff}}.
\label{eq:qsnr_fft_variance}
\end{equation}
Here $L_{\rm eff} = 2L+1-L(L+1)/M$ accounts for the finite number of measurement pairs contributing to the biased autocorrelation. Since the Fourier coefficients of the symmetric autocorrelation are real-valued, their absolute values follow a half-normal distribution rather than the Rayleigh distribution associated with the magnitude of complex Fourier coefficients \cite{byers-2014}.
\begin{equation}
p(\mathcal A_j)= \sqrt{\frac{2}{\pi V_j}} \exp\left(-\frac{\mathcal A_j^2}{2V_j}\right), \qquad \mathcal A_j\geq0.
\label{eq:qsnr_halfnormal}
\end{equation}
The standard deviation of the corresponding Fourier amplitude noise floor is therefore,
\begin{equation}
\sigma_{\mathcal A} =\sqrt{V_j\left(1-\frac{2}{\pi}\right)}
\simeq \frac{\sigma_{\rm{meas}}^2} {\sqrt{RK\,M L_{\rm eff}}} \sqrt{1-\frac{2}{\pi}}.
\label{eq:qsnr_noise_floor}
\end{equation}
Combining Eqs.~\eqref{eq:qsnr_peak} and \eqref{eq:qsnr_noise_floor}, we obtain the Fourier amplitude SNR,
\begin{equation}
\rm{SNR}_{\rm{QD}}\simeq\frac{A_Q^2\sqrt{RK\,ML_{\rm eff}}}{4\sigma_{\rm{meas}}^2\sqrt{1-2/\pi}}.
\label{eq:qsnr_final}
\end{equation}

\section{ SNR for CS and MCS}
\label{app:snr_cs_mcs}
We evaluate the spectral SNR of CS and MCS using the general framework developed in App.~\ref{app:snr}. In the weak-field regime, the two protocols share the same correlation amplitude, $A_{\rm{CS}}=A_{\rm{MCS}}=A_Q^2/2$, but differ in the temporal correlations of the random nuclear phase.
\subsection{Correlation spectroscopy}
For CS, the initial nuclear phase is independently randomized for each measurement. As in Eq.~\eqref{eq:cs_productsignal}, the measured signal is
\begin{equation}
Y_m^{(rk)}=A_{\rm{CS}}
\big[\cos\delta(\tau_s+\bar{\tau}_m) +\cos(2\epsilon_{\rm C,m}^{(rk)} +\delta(2\tau_s+\bar{\tau}_m))\big] +\xi_m^{(rk)}.
\label{eq:cs_snr_signal}
\end{equation}
Averaging over $R$ independent realizations and $K$ NV centers eliminates the phase-dependent term, giving
\begin{equation}
\langle Y_m\rangle =A_{\rm{CS}} \cos\delta(\tau_s+\bar{\tau}_m).
\label{eq:cs_snr_mean}
\end{equation}
The random nuclear phase, however, contributes to the measurement fluctuations as its variance . Since
$\epsilon_{\rm C,m}^{(rk)}$ is uniformly
distributed over $[0,2\pi]$, the variance of
the averaged signal is,
\begin{equation}
\sigma_{\rm{CS}}^2 =\frac{1}{RK} \left(\sigma_{\rm{meas}}^2 +\frac{A_{\rm{CS}}^2}{2}\right).
\label{eq:cs_snr_variance}
\end{equation}
These fluctuations are independent across sampling points under the assumed phase-randomization model. For $M$ uniformly sampled correlation delays, using the Gaussian spectral-noise result of App.~\ref{app:snr}, we obtain
\begin{equation}
\rm{SNR}_{\rm{CS}} \simeq \frac{A_{\rm{CS}}\sqrt{MRK}} {\sqrt{4-\pi} \sqrt{\sigma_{\rm{meas}}^2+ A_{\rm{CS}}^2/2}}.
\label{eq:cs_snr_final}
\end{equation}
\subsection{Multipoint correlation spectroscopy}
In MCS, the initial nuclear phase remains correlated throughout a complete measurement sequence. Using Eq.~(10), the measured response is,
\begin{equation}
Y_m^{(rk)}=A_{\rm{MCS}}\big[\cos(\delta mT_s)+\cos(2\epsilon_{\rm C}^{(rk)}+\delta mT_s+\delta\tau_s)\big] +\xi_m^{(rk)}.
\label{eq:mcs_snr_signal}
\end{equation}
Averaging over the independent nuclear phases yields
\begin{equation}
\langle Y_m\rangle
=A_{\rm{MCS}}\cos(\delta mT_s).
\label{eq:mcs_snr_mean}
\end{equation}
Unlike CS, the phase-dependent fluctuations are temporally correlated. Their covariance after averaging over $R$ realizations and $K$ NV centers is
\begin{equation}
\langle\eta_m\eta_{m'}\rangle=\frac{A_{\rm{MCS}}^2}{2RK} \cos[\delta(m-m')T_s] +\frac{\sigma_{\rm{meas}}^2}{RK} \delta_{mm'}.
\label{eq:mcs_snr_covariance}
\end{equation}
The first contribution is concentrated at $\delta$ and does not contribute to the spectral background when the signal is exactly aligned with a Fourier bin. Consequently, the background fluctuations are determined by the independent measurement noise, with effective variance $\sigma_Z^2=\sigma_{\rm{meas}}^2/(RK)$. Unlike CS, in which the initial nuclear phase is independently randomized at each sampling point, MCS preserves a common reference phase throughout the measurement trace. Consequently, the residual phase-dependent fluctuations are temporally correlated and contribute predominantly at the signal frequency, rather than forming an off-resonant spectral background. Applying the amplitude-SNR expression derived in App.~\ref{app:snr} gives,
\begin{equation}
\rm{SNR}_{\rm{MCS}} \simeq \frac{A_{\rm{MCS}}\sqrt{MRK}} {\sigma_{\rm{meas}}\sqrt{4-\pi}}.
\label{eq:mcs_snr_final}
\end{equation}
This expression uses the ensemble-averaged signal peak and the off-resonant spectral noise floor. For finite $RK$, the correlated nuclear-phase contribution also produces fluctuations of the signal peak itself, which are not included in the background-based SNR. Both expressions assume ideal correlation operations, negligible spectral leakage, and independent measurement noise.

\section{Spectral SNR for CASR}
\label{app:snr_casr}
We evaluate the CASR SNR using the general framework of App.~\ref{app:snr}, considering the Fourier power spectrum instead of the amplitude spectrum. Unlike Qdyne, CS, and MCS, CASR exploits coherent nuclear magnetization with a reproducible initial phase. The measured signal for the $r$th realization and $k$th NV center is
\begin{equation}
Y_m^{(rk)} =A_{\rm{CASR}}\cos(\delta mT_s+\phi_0) +\xi_m^{(rk)},
\label{eq:casr_snr_signal}
\end{equation}
Since $\phi_0$ is known, the signals can be averaged coherently over $R$ realizations and $K$ NV centers,
\begin{equation}
Z_m=\frac{1}{RK}\sum_{r,k}Y_m^{(rk)} =A_{\rm{CASR}}\cos(\delta mT_s+\phi_0) +\eta_m,
\label{eq:casr_snr_average}
\end{equation}
where the averaged noise satisfies $\langle\eta_m\rangle=0$ and $\langle\eta_m\eta_{m'}\rangle =\sigma_{\rm{meas}}^2\delta_{mm'}/(RK)$.
\end{appendices}
For $M$ uniformly sampled measurements, we compute the discrete Fourier transform using Eq.~\eqref{eq:snr_fft} and define the Fourier power spectrum as $\mathcal{P}_j=|\widetilde Z_j|^2$. When $\delta$ coincides with a nonzero, non-Nyquist Fourier bin, the signal contribution to the spectral peak is,
\begin{equation}
\mathcal{P}_{\rm{sig}} =\frac{A_{\rm{CASR}}^2}{4}.
\label{eq:casr_snr_peak}
\end{equation}
The Fourier noise coefficient at an off-resonant frequency can be written as $\widetilde\eta_j=X_j+iY_j$, where $X_j$ and $Y_j$ are independent, zero-mean Gaussian variables with variance $\langle X_j^2\rangle =\langle Y_j^2\rangle =\frac{\sigma_{\rm{meas}}^2}{2MRK}\equiv v$. Consequently, the Fourier noise power $\mathcal{P}_j=X_j^2+Y_j^2$ follows an exponential distribution \cite{Bendat2010},
\begin{equation}
p(\mathcal{P}_j) =\frac{1}{2v} \exp\left(-\frac{\mathcal{P}_j}{2v}\right), \qquad \mathcal{P}_j\geq0.
\label{eq:casr_snr_distribution}
\end{equation}
Its mean and variance are $\langle\mathcal{P}_j\rangle=2v$ and $\rm{Var}(\mathcal{P}_j)=4v^2$. Thus, the standard deviation of the Fourier power noise floor is,
\begin{equation}
\sigma_{\mathcal P} =2v=\frac{\sigma_{\rm{meas}}^2}{MRK}.
\label{eq:casr_snr_noise}
\end{equation}
Using the spectral SNR definition of
Eq.~\eqref{eq:snr_general}, with the power
spectrum as the spectral observable, we obtain
\begin{equation}
\rm{SNR}_{\rm{CASR}} =\frac{\mathcal{P}_{\rm{sig}}} {\sigma_{\mathcal P}} =\frac{A_{\rm{CASR}}^2MRK} {4\sigma_{\rm{meas}}^2}.
\label{eq:casr_snr_final}
\end{equation}
\end{document}